\documentclass{aa}
\usepackage{graphicx,natbib,amssymb,color}
\bibpunct{(}{)}{;}{a}{}{,}

\begin{document}

\title{The thermal instability of the warm absorber in NGC~3783}

\author{R. W. Goosmann \inst{1} \thanks{Email: rene.goosmann@astro.unistra.fr},
        T. Holczer \inst{2}, 
        M. Mouchet \inst{3}, 
        A.-M. Dumont \inst{3}, 
        E. Behar \inst{2},\\
        O. Godet \inst{4},
        A. C. Gon{\c c}alves \inst{1}, \and 
        S. Kaspi \inst{2}}

\institute{
$^1$~Observatoire Astronomique de Strasbourg, Universit\'e de Strasbourg, CNRS, UMR 7550, 11 rue de l’Universit\'e, F-67000 Strasbourg, France\\
$^2$~Department of Physics, Technion, Haifa IL-32000, Israel\\
$^3$~  LUTH, Observatoire de Paris, PSL Research University,  CNRS , Universit\'e Paris Diderot, Sorbonne Paris Cit\'e,  5 Place Jules Janssen, F-92195 Meudon, France\\
$^4$~Institut de Recherche en Astrophysique and Plan\'etologie, Universit\'e de Toulouse, CNRS, UMR 5277, 9 Avenue du Colonel Roche, F-31028 Toulouse Cedex 4, France\\
}

\offprints{Offprints are out of fashion!}

\authorrunning{Goosmann et al.}

\titlerunning{The thermal instability of the warm absorber in NGC~3783}

\abstract{The X-ray absorption spectra of active galactic nuclei frequently show evidence of winds with velocities in the order of $10^3$~km/s extending up to $10^4$~km/s in the case of ultra-fast outflows. At moderate velocities, these winds are often spectroscopically explained by assuming a number of absorbing clouds along the line of sight. In some cases it was shown that the absorbing clouds are in pressure equilibrium with each other.}
{We assume a photo-ionized medium with a uniform total (gas+radiation) pressure. The irradiation causes the wind to be radiation pressure compressed (RPC). We attempt to reproduce the observed spectral continuum shape, ionic column densities, and X-ray absorption measure distribution (AMD) of the extensively observed warm absorber in the Seyfert galaxy NGC~3783.}
{We compare the observational characteristics derived from the 900~ks Chandra observation to radiative transfer computations in pressure equilibrium using the radiative transfer code {\sc titan}. We explore different values of the ionization parameter $\xi$ of the incident flux and adjust the hydrogen-equivalent column density, $N_{\rm H}^0$, of the warm absorber to match the observed soft X-ray continuum. From the resulting models we derive the column densities for a broad range of ionic species of iron and neon and a theoretical AMD that we compare to the observations.}
{We find an extension of the degeneracy between $\xi$ and $N_{\rm H}^0$ for the constant pressure models previously discussed for NGC~3783. Including the ionic column densities of iron and neon in the comparison between observations and data we conclude that a range of ionization parameters between 4000 and 8000~ergs~cm~s$^{-1}$ is preferred. For the first time, we present theoretical AMD for a constant pressure wind in NGC~3783 that correctly reproduces the observed level and is in approximate agreement with the observational appearance of an instability region.}
{Using a variety of observational indicators, we confirm that the X-ray outflow of NGC~3783 can be described as an RPC medium in pressure equilibrium. The observed AMD agrees with a uniformly hot or a uniformly cold thermal state. The measured ionic column densities suggest that the wind tends to the uniformly cold thermal state. The occurrence of thermal instability in the warm absorber model may depend on the computational method and the spatial scale on which the radiative transfer is solved.}
  

\date{Received 22 October 2014; accepted 25 February 2016}

\maketitle

\section{Introduction}
\label{sect:intro}

The presence of outflows in active galactic nuclei (AGN) shows that black hole accretion is not a one-way process. Part of the matter pulled in by the black hole is not accreted but re-ejected into ballistic jets or non-collimated winds. Studying the geometry and dynamics of AGN winds is important for understanding their launch from an accretion disk \citep{fukumura2014,higginbottom2014}. Furthermore, it has become clear that winds contribute to the mechanical AGN feedback onto the host galaxy \citep{king2005,crenshaw2012,tombesi2013} and may influence star formation in a complex way \citep{nayakshin2012}. Finally, if the scaling of black hole properties is valid over a broad range of masses, the physics of AGN winds also has implications for the study of black hole X-ray binaries \citep{king2013}.

Non-collimated outflows in AGN are identified by shifted and broadened absorption edges or lines in the X-ray and UV spectrum of many Seyfert-1 galaxies and quasars \citep[see, e.g.,][]{halpern1984,nandra1994,komossa1997,reynolds1997,crenshaw1999,wang1999,hamann2000,monier2001,matsumoto2004,ebrero2013}. When detected in the X-ray band the absorbing wind is called ``warm absorber'' and attributed to a photo-ionized medium seen mostly in absorption with a relatively weak component of reprocessed emission.

Our understanding of the composition and dynamics of AGN winds has greatly improved by using high resolution gratings aboard large X-ray observatories, such as the {\it Low and High Energy Transmission Grating Spectrometers} of {\it Chandra} \citep[see, e.g.,][]{kaastra2000,kaspi2000,lee2001,steenbrugge2003a,fiore2003,fang2005,mckernan2007,steenbrugge2009,reeves2013} or the {\it Reflection Grating Spectrometer} of {\it XMM-Newton} \citep[see, e.g.,][]{blustin2002,mason2003,steenbrugge2003b,ashton2004,schurch2006,krongold2009,longinotti2010,ricci2010,mehdipour2012}. Multi-waveband campaigns provided the detailed broad-band spectrum and the time-dependent behavior of X-ray outflows in a number of bright Seyfert-1 galaxies: NGC~7469 \citep{blustin2003}, NGC~5548 \citep{crenshaw2003,kaastra2004,steenbrugge2005}, Mrk~509 \citep{detmers2011,steenbrugge2011,kaastra2011,ebrero2011,kaastra2012,arav2012}, IRAS~13449+2438 \citep{lee2013,digesu2013,digesu2014}, NGC~3783 \citep{kaspi2002,gabel2003,netzer2003}.

With the improved quality of the observations came substantial improvements in the spectral modeling. Advanced X-ray radiative transfer codes were combined with state-of-the-art atomic data bases to analyze the high-resolution data. Numerical codes like {\sc cloudy} \citep{ferland1998,ferland2013} and {\sc xstar} \citep{kallman2001,bautista2001,garcia2005,garcia2009} are frequently used to model the conditions of the photo-ionization equilibrium, the intensity and spectral shape of the incident radiation, and the total hydrogen-equivalent column density of the warm absorber with its numerous absorption and emission lines in the soft X-ray range. In this work, we apply the code {\sc titan} initially presented in \citet{dumont2000} and updated in \citet{dumont2003,collin2004,goncalves2007}. It was used to pioneer the modeling of a warm absorber in pressure equilibrium \citep{rozanska2006,goncalves2006,goncalves2007}.

The progress in X-ray spectroscopy and modeling over the last 15 years revealed many details about AGN winds. Nonetheless, for a complete understanding some key elements still need to be found, such as the structure and the thermal state of the wind. In this work, we further explore the hypothesis that the X-ray outflow in the Seyfert galaxy NGC~3783 is a continuous, photo-ionized medium in total pressure equilibrium. Our radiative transfer modeling focuses in particular on the reproduction of the observed ionic column densities, the total hydrogen-equivalent column density, and the absorption measure distribution (AMD) of the warm absorber \citep{holczer2007}.

The paper is organized as follows: in Sect.~\ref{sec:ngc3783}, we recall previous observational and modeling work on the warm absorber of NGC~3783. The following Sect.~\ref{sec:AMD_update} presents an updated version of the AMD analysis and points out the adjustments with respect to the previous study. Our modeling approach is described in Sect.~\ref{sec:model} and the results are presented in Sect~\ref{sec:results}. We further discuss our work in Sect.~\ref{sec:discuss} before drawing our conclusions in Sect.~\ref{sec:conclude}.


\section{A warm absorber in pressure equilibrium}
\label{sec:ngc3783}

The warm absorber of the X-ray bright narrow-line Seyfert-1 galaxy NGC~3783 was extensively observed during a multi-wavelength campaign that involved a 900~ks observation with {\it Chandra} \citep{kaspi2001} and 280~ks of {\it XMM-Newton} time \citep{behar2003}. The total {\it Chandra} observation was collected over six periods of $\sim 170$~ks each and gave rise to a warm absorber spectrum with very high signal-to-noise. A detailed spectral analysis was carried out for the time-integrated spectrum \citep{kaspi2002} and for individual observation periods \citep{netzer2003}. It turned out that on time scales larger than $\sim 100$~ks the source switches between two different flux states. The shape of the spectral model in both states is very similar suggesting that these flux variations are intrinsic to the source and do not significantly alter the structure of the warm absorber. The warm absorber could be divided into two systematic velocity parts that were in agreement with the velocity structure of the associated UV absorber \citep{gabel2003}. Each velocity component contains three spectroscopic components with high, intermediate and low ionization states \citep{kaspi2001,kaspi2002}. The three spectral components could be modeled by a combination of constant density clouds irradiated from the backside by the same incident spectrum. Using a different modeling method, \citet{krongold2003} found similar results but only assuming two spectroscopic components.

In the following, it turned out that the individual spectroscopic components of the warm absorber in NGC~3783 lie at the same ratio of radiation-to-gas pressure \citep{krolik2001,netzer2003,krongold2003}. Thereby, the radiation pressure largely dominates compared to the gas pressure. This suggests that the two or three absorbing clouds are confined by the radiation pressure, a model recently explored by \citet{stern2014a,stern2014b} and \citet{baskin2014} for the line emitting regions in active galactic nuclei. A radiation pressure compressed (RPC) wind can be realized by assuming a photo-ionized medium that is isobaric in terms of the total (gas+radiation) pressure. Such models were explored in \citet{rozanska2006} and \citet{goncalves2007}. For the case of NGC~3783, \citet{goncalves2006} presented spectral modeling of an isobaric RPC wind and reproduced rather accurately the continuum shape of the soft X-ray spectrum. The continuous medium turned out to be divided into three ionization states: thermal instability induced two steep temperature drops between the high/intermediate and intermediate/low temperature zones. Such a structure is also compatible with the UV observations of the low-ionization absorbers in NGC~3783 \citep{gabel2003}.

The thermal instability of photo-ionized media was first studied in the AGN context by \citet{krolik1981}. A key figure to visualize the instability is the ``S-shaped'' curve tracing the temperature of the medium as a function of the pressure ratio $P_{\rm rad}/P_{\rm gas}$. Such curves have been studied in detail for photo-ionized media at constant density and in the optically thin limit \citep{hess1997,krolik1995,krolik2001,chakravorty2008,chakravorty2009,chakravorty2012}. All these modeling campaigns revealed a complex but systematic dependence of the thermal instability on the shape of the incident spectrum and the element abundances. \Citet{goncalves2007} produced theoretical instability curves for a photo-ionized RPC medium and analyzed the dependency of the ``S-curve'' on different slopes of the intrinsic spectrum and on the total column density of the warm absorber. The study concludes that the thermal instability is a robust phenomenon occurring over a broad range of incident spectra and column densities of the RPC wind.

For the warm absorber of NGC~3783, \citet{goncalves2006} found a spectral degeneracy between the intensity of the incident spectrum and the column density of the warm absorber. The degeneracy makes it difficult to constrain the parameters of the wind. In the present work, we therefore add the observed column densities and the AMD to the comparison and we explore a wider parameter range to test if the degeneracy can be removed.


\section{Updating the soft X-ray AMD}
\label{sec:AMD_update}

The soft X-ray spectrum of an AGN outflow allows us to derive the column density of individual ionic species by measuring the equivalent width of their absorption lines. It was shown by \citet{steenbrugge2003b} for the case of the Seyfert galaxy NGC~5548 that the column density of different ionic species in the warm absorber can cover a broad range in ionization parameter (counting more than three orders of magnitude). The derivative of the cumulative hydrogen column density in ionization parameter $dN_{\rm H}/d\xi$ can be related to the measured column density of the individual ionic species establishing a link between the ionization profile of the medium and observables of the warm absorber. The approach was further developed by \citet{holczer2007} who introduced the AMD and included physical relations between the ionization parameter $\xi$ and the fractional ionic abundances, $f_{\rm ion}$, at a given location inside the wind. The AMD is given as: 

\begin{equation}
  {\rm AMD} = \frac{dN_{\rm H}}{d\log\xi}, \;
  N_{\rm H} = \int {\rm AMD} \, d\log\xi.
  \label{eqn:amd}
\end{equation}

Herein, $N_{\rm H}$ is the cumulative, hydrogen-equivalent column density at a given position inside the medium and the ionization parameter is defined by 

\begin{equation}
  \xi = \frac{4\pi}{n_{\rm H}}\int_{1 \, \rm Ryd}^{1000 \, \rm Ryd} \,F_{\rm ion} \, d\nu
  \label{eqn:xi_obs}
\end{equation}

with $F_{\rm ion}$ being the ionizing flux and $n_{\rm H}$ the hydrogen number density. Assuming a set of chemical abundances $A_{\rm Z}$ the ionic column densities are expressed by:

\begin{equation}
  N_{\rm ion} = A_{\rm Z} \int {\rm AMD} \, f_{\rm ion}(\xi) \, d\log\xi.
\end{equation}

\begin{figure}[!t]
  \centering
  \includegraphics[width=0.4975\textwidth]{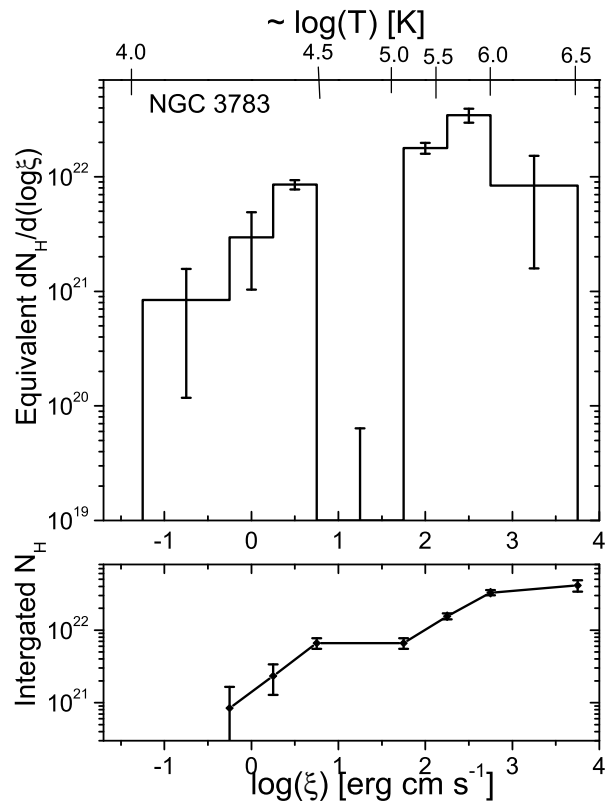}
  \caption{Absorption measure distribution for the warm absorber of NGC~3783 using the solar chemical abundances measured by \citet{asplund2009}. The top panel shows the differential form of the AMD. In the bottom panel it is integrated up to the total column density (see equation (\ref{eqn:amd})). A scale in $\xi$ and in temperature $T$ are provided with the relation $T(\xi)$ being obtained from the {\sc xstar} computations of the fractional ionic abundances (see text).}
  \label{fig:amd}
\end{figure}

With the set of $N_{\rm ion}$ being determined from the absorption spectrum, the AMD can be computed if the relationships between $f_{\rm ion}$ and $\xi$ are known. They were obtained applying the radiative transfer code {\sc xstar 2.1kn3} \citep{kallman2001} for a range of $\xi$ parameters and a set of chemical abundances $A_{\rm Z}$. The incident spectrum was modeled as a broken power law \citep[see Fig. 7 in][]{kaspi2002}. \Citet{holczer2007} computed the AMD of the warm absorber in NGC~3783 assuming the solar abundances reported by \citet{asplund2005}. These abundances were updated in \citet{asplund2009} revealing a few significant differences with respect to the earlier measurements. We therefore plot here the recomputed AMD for NGC~3783 in Fig.~\ref{fig:amd} \citep[see also][]{holczer2012}.

Every ionic column density $N_{\rm ion}$ is determined from a set of several absorption lines. To have a sufficiently large coverage of the AMD as a function of $\xi$, it is important to measure lines that all together represent a wide range of ionization states. The iron absorption lines cover ionization states starting from Fe~{\scriptsize IV} up to Fe~{\scriptsize XXVI}, which is why iron is a particularly important element in the AMD analysis. Apart from iron, the AMD in Fig.~\ref{fig:amd} compiles a set of absorption lines from Ne~{\scriptsize VII} up to Ne~{\scriptsize X}. Since the distribution of absorption lines in $\xi$ is finite and discrete, the resulting AMD is of limited resolution and thus approximated by a step function. The associated error bars are obtained from a Monte-Carlo fitting method and also comprise the errors associated to the measurement of the equivalent line widths \citep[see][for more details on the method]{holczer2007}.

Compared to the earlier version of the AMD shown in \citet{holczer2007} the new element abundances of \citet{asplund2009} do not cause any major changes. As before, a ``forbidden range'' in temperature is detected across which the AMD drops to zero. This interval in ionization parameter and temperature is presumably associated to a thermal instability. In the following, we explore this conjecture by modeling coherently the radiative transfer inside the warm absorber.


\section{Radiative transfer modeling with {\sc titan}}
\label{sec:model}

Our modeling method is similar to the one adopted in \citet{goncalves2006} and involves the radiative transfer code {\sc titan} \citep{dumont2000,dumont2003,collin2004,goncalves2007}. The geometrical depth inside the infinitely wide, plane-parallel medium is measured along the coordinate axis $z$ and the slab is irradiated at $z = 0$ by an incident spectrum, $I_{\rm inc}(E)$. The angular distribution of the irradiation is closely collimated to the line of sight along $z$. At $z=z_{\rm tot}$, the observed spectrum emerges.  Between the two limiting surfaces, the medium is divided into a few hundreds of layers and for each layer the radiative transfer is solved under the condition that the total pressure, $P_{\rm tot} = P_{\rm gas} + P_{\rm rad}$, remains constant throughout the medium. Our calculations include all relevant absorption, re-emission and scattering effects. The profiles for temperature, fractional ionized abundances and density are computed. The code coherently computes the radiative transfer along the line of sight, i.e., along the $z$-axis at $\mu = \cos i = 1$ and at five additional inclinations $i$ chosen to optimize angular integration. We thus add to the pure absorption spectrum seen along the line of sight the re-emission from the slab. The radiative transfer equation is solved along six directions using an {\it Accelerated Lambda Iteration} method and assuming a stationary state for all atomic processes such as photo-ionization and recombination or line excitation and de-excitation. The relevant heating and cooling processes are taken into account. More details about the code and its successive developments can be found in the aforementioned four {\sc titan} papers.

The presence of thermal instability at a given depth inside the medium complicates the radiative transfer due to very steep temperature drops. In addition to this, for the same local incident flux two or more stable temperature solutions are possible. For this reason, several physical states of the medium may coexist at the same time, which leads to a dynamical fragmentation of the gas. One can imagine, for instance, that relatively cooler and denser gas clouds are embedded in a hotter, more dilute ambient medium \citep{krolik1995}. Such a thermal structure would certainly be time-variable; different regions of the cloud are expected to be in different thermal states and changes between those states are supposed to happen on time scales limited by the dynamical time scale:

\begin{equation}
  t_{\rm dyn} \sim 2 \times 10^4 \frac{f_{\rm inst}N_{23}}{\sqrt{T_5}n_{12}} \; [s].
\end{equation}

Herein, $n_{12}$ denotes the volume density in units of $10^{12}$~cm$^{-3}$, $T_5$ the temperature of the medium in units of $10^5$~K, and $N_{23}$ the total column density in units of $10^{23}$~cm$^{-2}$. The value $f_{\rm inst}$ denotes the fraction of the total column density that is in the instability regime. For a detailed discussion of relevant time scales in the warm absorber we refer to \citet{goncalves2007} and \citet{chevallier2007}. In the parameter range that we investigate here we can estimate $t_{\rm dyn}$ to be less than a few 10$^4$~s, which is within the X-ray exposure times considered.

From the preceding estimation it follows that a complete, time-dependent treatment of the dynamical and thermal structure together with the radiative transfer inside the warm absorber in NGC~3783 is not necessary. We also note that our radiative transfer calculations do not include any dynamics of the medium on larger spatial scales. The relative motion between different parts of the wind are thus neglected, which is a reasonable simplification for the study of NGC~3783 since its X-ray spectrum shows rather low values of line blueshifts. This was shown in a recent study with {\it Chandra} that did not reveal any significant differences in the dynamics of the X-ray absorber in NGC~3783 when compared to the epoch of 2001/2002 \citep{scott2014}.

The thermal instability can be included in the model using different computational modes of {\sc titan}. It is possible to assume a {\it uniquely cold} solution, a {\it uniquely hot} solution or an {\it approximate intermediate} solution for the temperature profile along $z$. The latter computational mode suffers from the fact that in a stationary scheme it cannot be computed in a completely self-consistent way. The uniquely hot and cold solutions, on the other hand, represent the two most extreme thermal states the medium can adopt. If one chooses the uniformly hot (cold) computation mode, {\sc titan} assumes at every thermally unstable position $z$ the hottest (coldest) stable solution for the temperature balance. These solutions are thus entirely physical and set the limits in between which the medium as a whole can thermally evolve. The intermediate temperature solution, approximation for the radiative coupling, represents a spatial average across the multiple-phase medium while respecting a coherent solution of the radiative transfer. For more details about the different calculation modes of {\sc titan}, we refer to \citet{goncalves2007} and references therein.

The temperature solution at a given position $z_{\rm i}$ is found iteratively by adjusting the local density $n_{\rm i}$ such that the total pressure remains constant and in agreement with the incoming and outgoing radiative energy going to or coming from the layers $z_{\rm i-1}$ and $z_{\rm i+1}$. At the limits $z = 0$ and $z = z_{\rm tot}$ the incoming and outgoing radiation in the forward direction are identical with the incident and the observed spectrum, respectively. Along the line of sight, the local spectrum evolves due to absorption, scattering and re-emission processes. Therefore, the local temperature solution in each layer $z_{\rm i}$ of the medium is connected to the overall solution of the radiation field. The presence of thermal instability leads to sharp drops in the temperature structure, which causes numerical difficulties in solving the radiative transfer equation. The $z$-grid needs a very fine stratification across the sharp temperature drops, which is constantly adjusted during the computation and unavoidably increases the necessary CPU time.

The latest version of {\sc titan} includes several improvements with respect to our previous work  \citep{goncalves2006} . The number of spectral lines included in the computations was raised significantly and now 4000 line transitions are taken into account, in particular the unresolved transition arrays (UTA) of various ionization states of iron. We adopt the element abundances of H, He, C, O, N, Ne, Mg, S, Si, and Fe as given by \citet{asplund2009} and used in the AMD analysis of Sect.~\ref{sec:AMD_update}. We adopt the definition of the (numerical) ionization parameter $\xi_{\rm tot}$ defined by

\begin{equation}
  \xi_{\rm tot} = \frac{4\pi F_{\rm tot}}{n_{\rm H}}
\end{equation}

where $n_{\rm H}$ denotes the hydrogen volume density at the irradiated surface and $F_{\rm tot}$ the integrated incident flux:

\begin{equation}
  F_{\rm tot} = \int_0^\infty F_{\rm inc}(\nu) {\rm d}\nu
\end{equation}

The above definition of $\xi$ differs from the one given in equation (\ref{eqn:xi_obs}) by integration over an infinite energy scale. The total flux $F_{\rm tot}$ depends on the spectral shape across the UV and X-ray range. We have adopted here the incident spectrum that was introduced and justified by the modeling work of \citet[][see their figure~7]{kaspi2001}. Such an incident spectrum was also adopted by \citet{goncalves2006}. We mention here that in the more general analysis of \citet{goncalves2007}, the incident spectrum had a power law shape with a uniform photon index $\Gamma = 2$ across the whole spectral range.  With the incident continuum adopted for our modeling the difference between both definitions of $\xi$ amounts to about 20\%. In our modeling results we take this conversion into account.

Turbulence inside the medium can alter the radiative transfer, especially the shape of absorption lines. For a fragmented warm absorber, we distinguish between micro- and macro-turbulence: the micro-turbulence refers to a turbulent velocity distribution inside each individual cloud and therefore adds to the thermal line broadening. Therefore, it can have an impact on the strength of resonant scattering lines. The macro-turbulence is due to the velocity distribution of the individual absorbing clumps all together. Assuming that the covering factor of the clouds is small, photons emerging from one cloud are unlikely to further interact with the medium. Therefore the macroscopic turbulence does not have the same effect on the radiative transfer in spectral lines. For more details about the two types of turbulence we refer to \citet{godet2004}. In this modeling, we suppose that only micro-turbulence is relevant. The velocities derived from various blueshifted absorption lines do not depend on the ionization potential of the lines with a mean velocity of -590~km/s and a small dispersion of 150~km/s \citep{kaspi2000}. The use of the code {\sc titan} that cannot include the internal dynamics of the medium is thus justified in the absence of strong velocity gradients.

Our primary goal is to compare the observed ionic column densities and the related AMD to a set of {\sc titan} models obtained under the assumption of pressure equilibrium. We compute models for five values of the ionization parameter between $\xi_{\rm tot} = 2000$~ergs~cm~s$^{-1}$ and $\xi_{\rm tot} = 20000$~ergs~cm~s$^{-1}$. For a given $\xi_{\rm tot}$, we explore the two extreme thermal states, the ``cold'' and the ``hot'' solutions. For one case (hot solution of $\xi_{\rm tot} = 8000$~ergs~cm~s$^{-1}$) we also investigate the effects of the turbulent velocity using a higher value of 400 km/s. Apart from $\xi_{\rm tot}$ and the thermal state of the medium, all model ingredients are fixed by the observational analysis: element abundances, turbulent velocity, and incident spectral shape.


\section{Modeling results}
\label{sec:results}

We adjust the total column density $N_{\rm H}^0$ of the medium until the observed spectral continuum is correctly reproduced. For this purpose, we use the softness ratio taken between the two wavelength bands 15--25~\AA ~ and 2--10~\AA~as introduced by \citet{netzer2003}.

\begin{figure}[t!]
  \includegraphics[width=0.3775\textwidth,angle=-90]{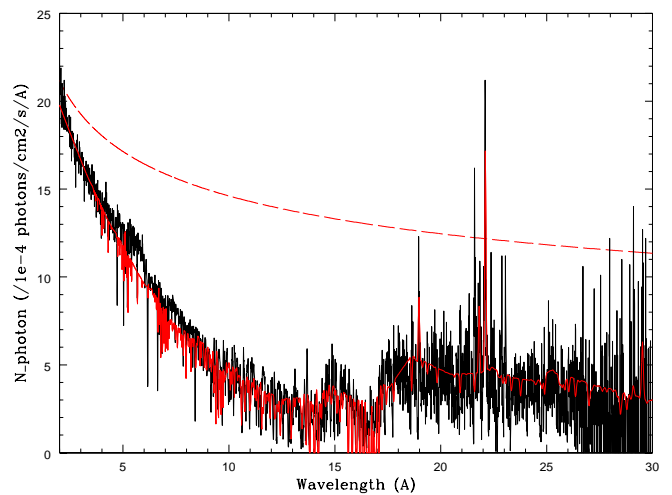}  
  \caption{Comparison of the modeled (red full curve) to the observed (black curve)  spectral continuum shape for the hot model case of $\xi_{\rm tot} = 8000$ ergs~cm~s$^{-1}$ and $N_{\rm H}^0 = 7.38\times 10^{22}$~cm$^{-2}$.  The dashed line is the incident spectrum.
  \label{fig:comp_spe_xi8e3}}
\end{figure}

For a given $\xi_{\rm tot}$, this softness ratio decreases when the total column density increases. We keep the value $N_{\rm H}^0$ for which this ratio is the closest to the observed one of 0.093 as derived from the {\it Chandra} data. We show a spectral comparison between the data and an exemplary model with $\xi_{\rm tot} = 8000$~ergs~cm~s$^{-1}$ in Fig.~\ref{fig:comp_spe_xi8e3} for the hot solution.

For each pair ($\xi_{\rm tot}$, $N_{\rm H}^0$) obtained in this way (Sect.~\ref{sec:tot_NH}), we compare the observed ionic column densities to the model values (Sect.~\ref{sec:comp_Ni}) and we investigate the model temperature profiles (Sect.~\ref{sec:comp_T}) as well as the model AMD (Sect.~\ref{sec:AMD}) across the warm absorber.

\subsection{The total hydrogen column density}
\label{sec:tot_NH}

The modeled hydrogen column density, $N_{\rm H}^0$, of the warm absorber as a function of the ionization parameter and the thermal state of the medium is summarized in Table~\ref{tab:xi_NH}. 

\begin{table}
 \centering
 \caption{Total column density $N_{\rm H}^0$ (in~cm$^{-2}$) of the warm absorber medium as a function of the theoretical ionization parameter $\xi_{\rm tot}$ (in ergs~cm~s$^{-1}$) and the thermal state.}
 \label{tab:xi_NH} 
 \begin{tabular}{ccccc}
  \hline\hline
     ~             &  {\bf cold state}   &   {\bf hot state}  \\
     $\xi_{\rm tot}$ &    $N_{\rm H}^0$     &    $N_{\rm H}^0$     \\
  \hline
     2000 & $2.3\times 10^{22}$  & $3.2\times 10^{22}$\\
     4000 & $4.7\times 10^{22}$  & $6.2\times 10^{22}$\\
     8000 & $6.0\times 10^{22}$  & $7.4\times 10^{22}$\\
    16000 & $7.6\times 10^{22}$  & $8.9\times 10^{22}$\\
    20000 & $8.3\times 10^{22}$  & $9.0\times 10^{22}$\\ 
  \hline\hline
 \end{tabular}
\end{table}

The resulting $N_{\rm H}^0$ increases with $\xi_{\rm tot}$, which is expected. For a stronger ionizing flux the high temperature layer on the irradiated side of the medium is thicker and therefore a higher total column density is necessary to obtain lower temperature regions on the far side. For the same reason the $N_{\rm H}^0$ obtained at a given $\xi_{\rm tot}$ is higher for the hot solution than for the cold solution. Indeed, the computational mode of always choosing the cold over the hot stable solution when the medium is unstable produces low temperature regions at smaller distances from the irradiated surface of the medium.

The observed total column density of $4 \times 10^{22}$~cm$^{-2}$ derived from the AMD analysis (see bottom panel of Fig.~\ref{fig:amd}) is compatible with a low $\xi_{\rm tot}$ value between 2000 and 4000~ergs~cm~s$^{-1}$ for the cold as well as for the hot solution. Note, however, that only very few lines are formed in the hottest part of the column where the material is almost fully ionized. In this temperature range, the estimation of the total column density depends strongly on only a few absorption lines of highly ionized iron. We show in the following that these observed ionic column densities have large error bars contributing to a significant uncertainty of the observed total column density.

\Citet{goncalves2006} derived a best solution with $\xi_{\rm tot}= 2500$ ergs~cm~s$^{-1}$ and $N_{\rm H}^0 = 4 \times 10^{22}$~cm$^{-2}$, which is marginally consistent with the range of best solutions found in this analysis and in agreement with the column density derived from the observed AMD. Their solution was obtained by visually finding the best possible match between the continuum model and the observed spectrum. We note that the range of models examined by \citet{goncalves2006} was narrower and, more importantly, we recall that at the time only ``intermediate'' solutions were considered. Futhermore, the number of line transitions included in the code was much lower (900 versus 4000). Nevertheless the total $N_{\rm H}^0$ derived by \citet{goncalves2006} is of the same order of magnitude as the values we derive for the solutions at lower $\xi_{\rm tot}$. Moreover, as already mentioned by  \citet{goncalves2006}, the spectrum can be fit by degenerate pairs of solutions. This degeneracy can be partly removed when comparing the ionic column densities of the models to those obtained from the observed spectrum (see Section 5.2). In \citet{goncalves2006} this comparison was done using the measurements by \citet{netzer2003} that did not include iron species. We show below that high ionization states are not reproduced in a satisfactory manner by low $\xi_{\rm tot}$ models, and thus prevented \citet{goncalves2006} from favoring a solution with higher $\xi_{\rm tot}$. The results in Table~1 were obtained with the same turbulent velocity (150 km/s) as in \citet{goncalves2006}. Here, the hot solution of the $\xi_{\rm tot}= 8000$ model was also computed with a turbulent velocity of 400 km/s and gave a total column density of $7.0\,10^{22}$~cm$^{-2}$. This is only $5\%$ less than for a turbulent velocity of 150 km/s.

\subsection{The column density of individual ionic species}
\label{sec:comp_Ni}

\begin{figure*}[!th]
  \includegraphics[angle=-90,width=0.4975\textwidth,clip]{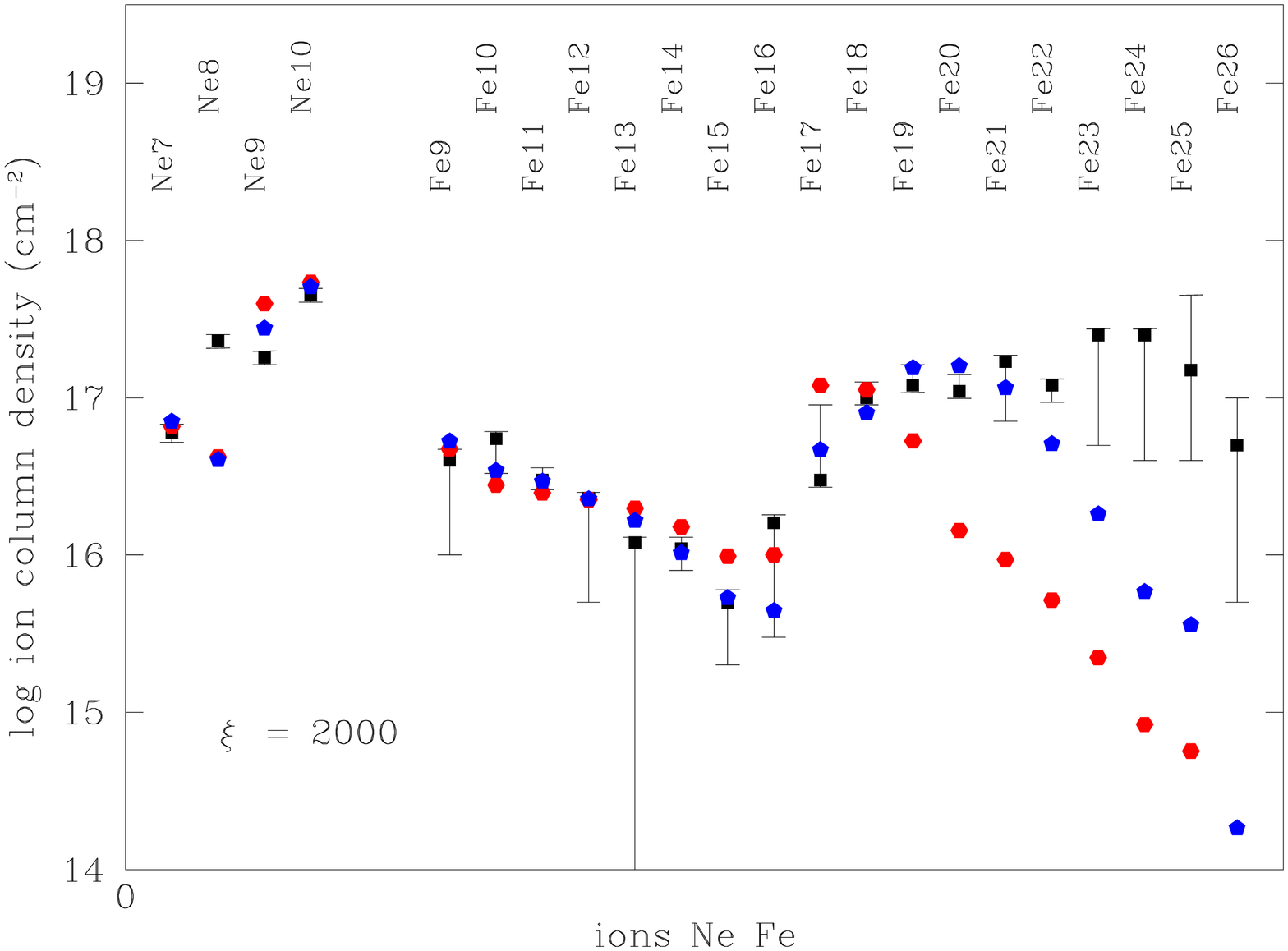}
  \includegraphics[angle=-90,width=0.4975\textwidth,clip]{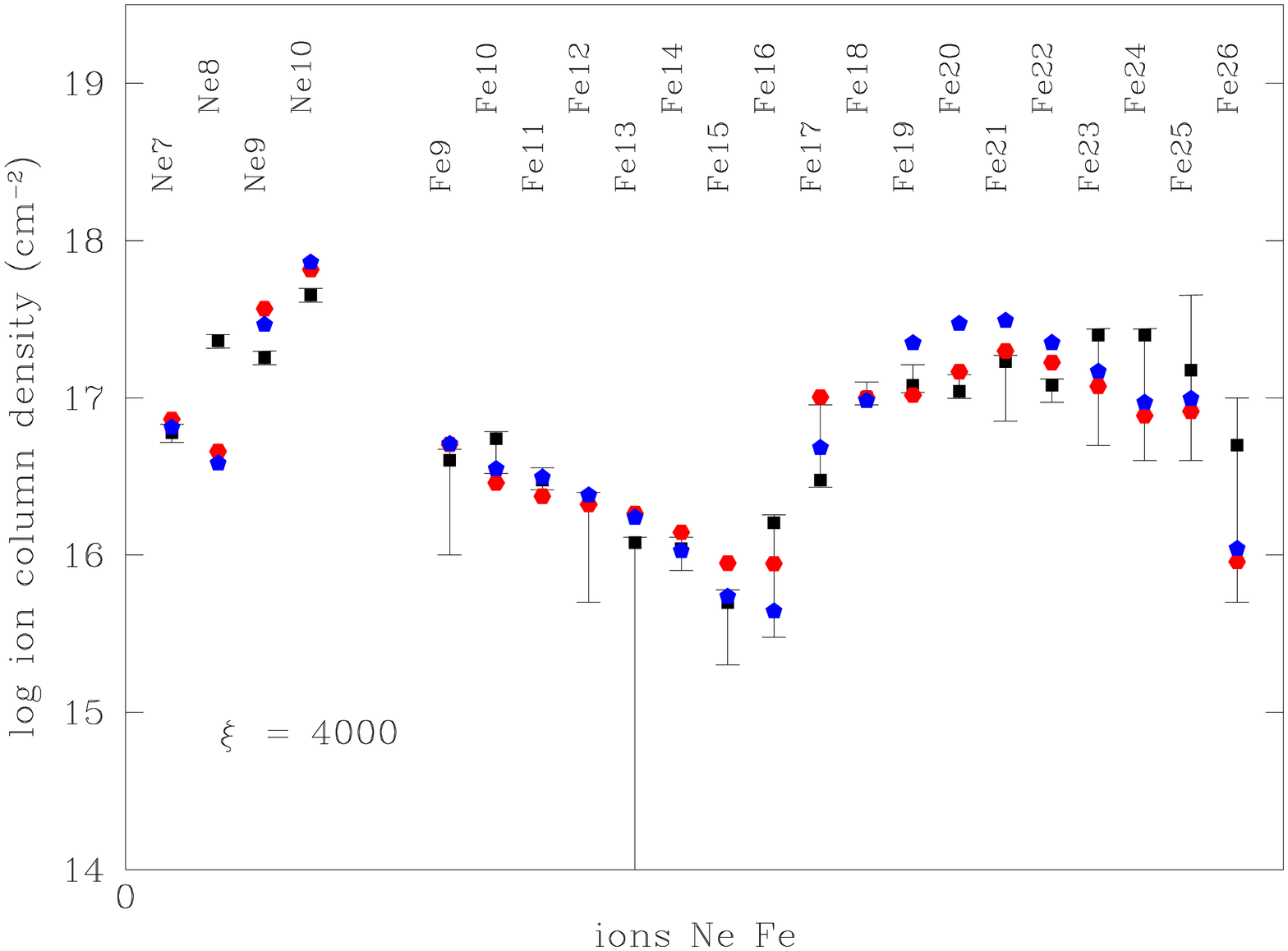}\\
  \includegraphics[angle=-90,width=0.4975\textwidth,clip]{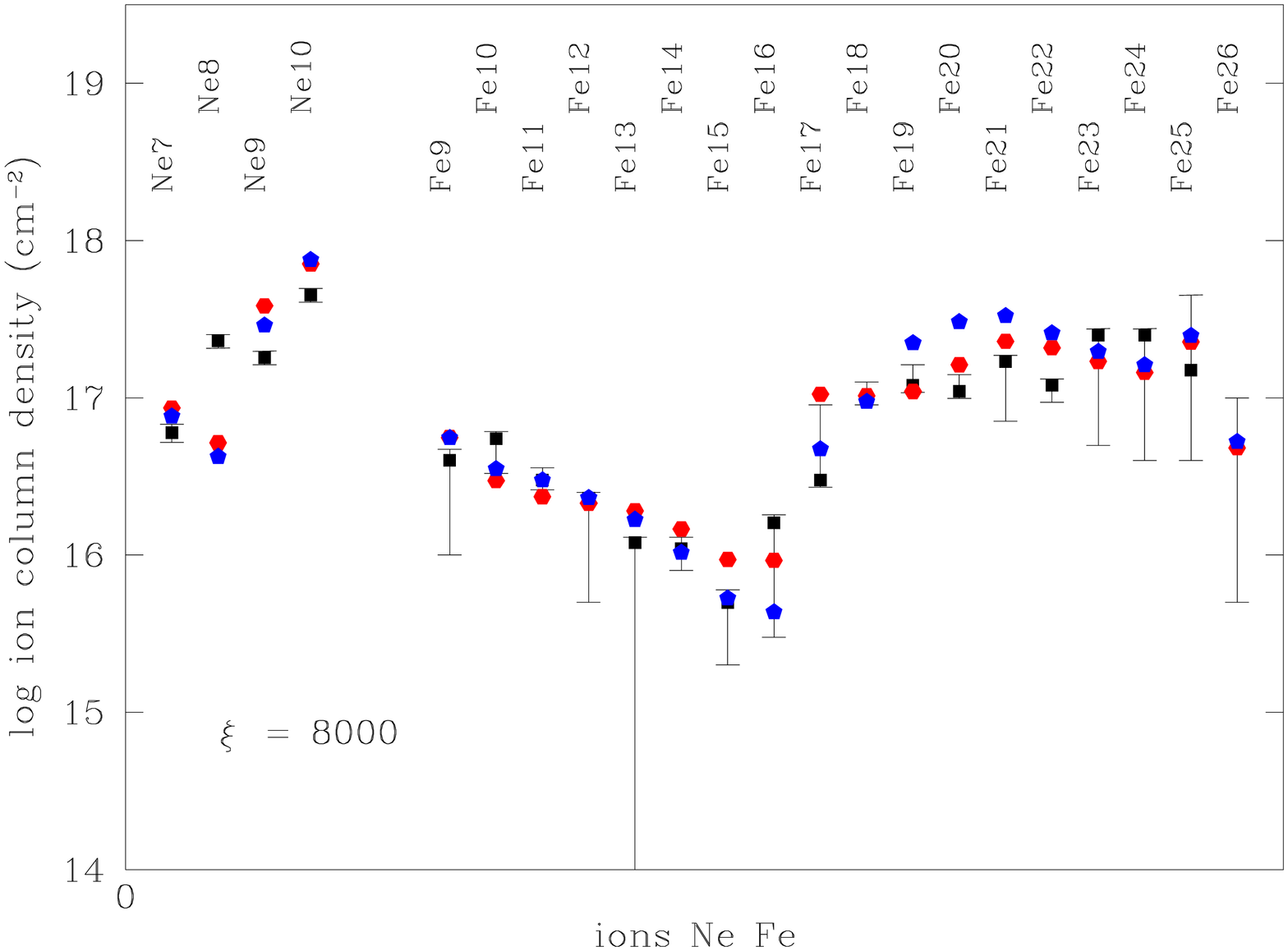}
  \includegraphics[angle=-90,width=0.4975\textwidth,clip]{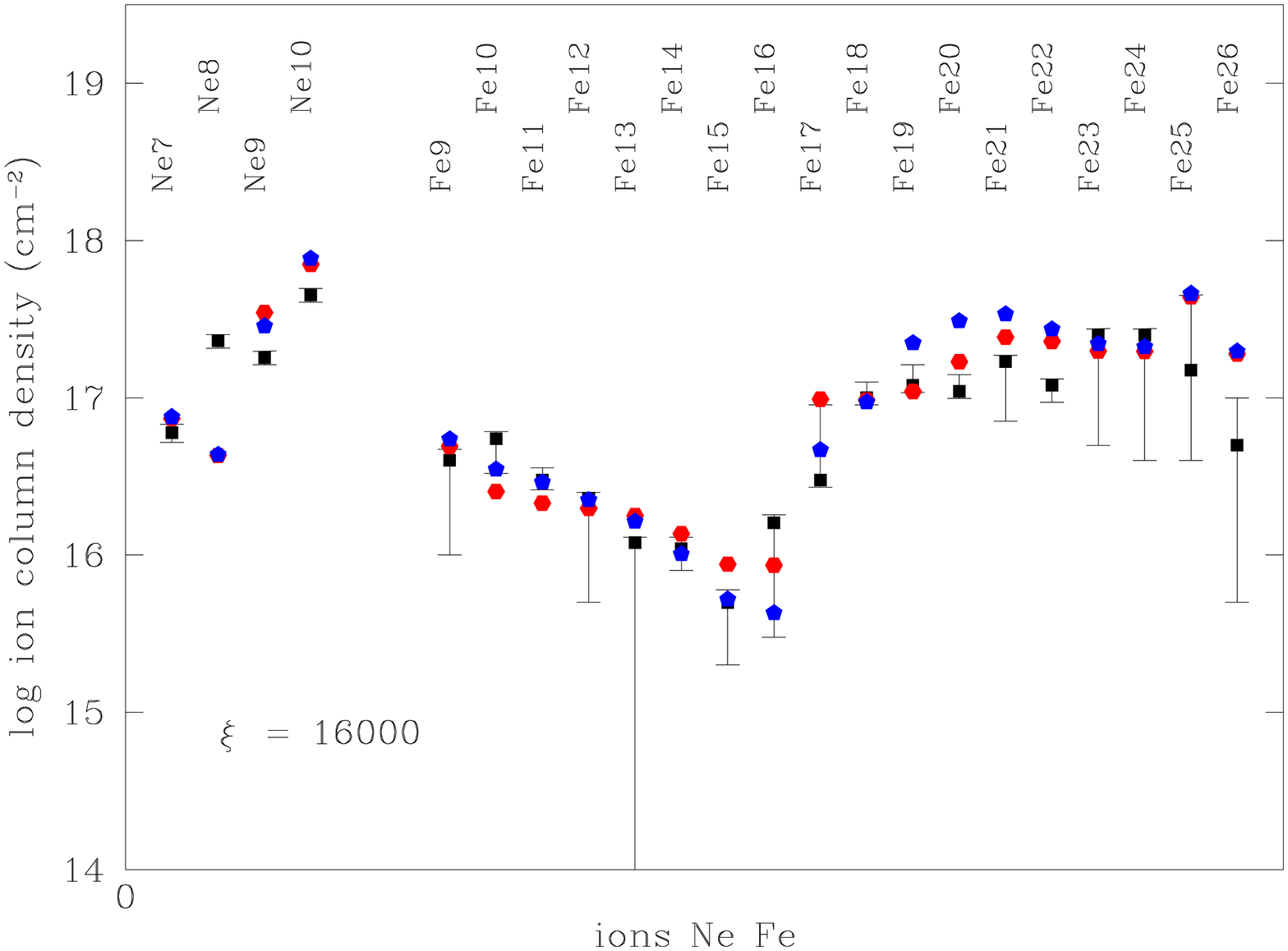}\\
  \includegraphics[angle=-90,width=0.4975\textwidth,clip]{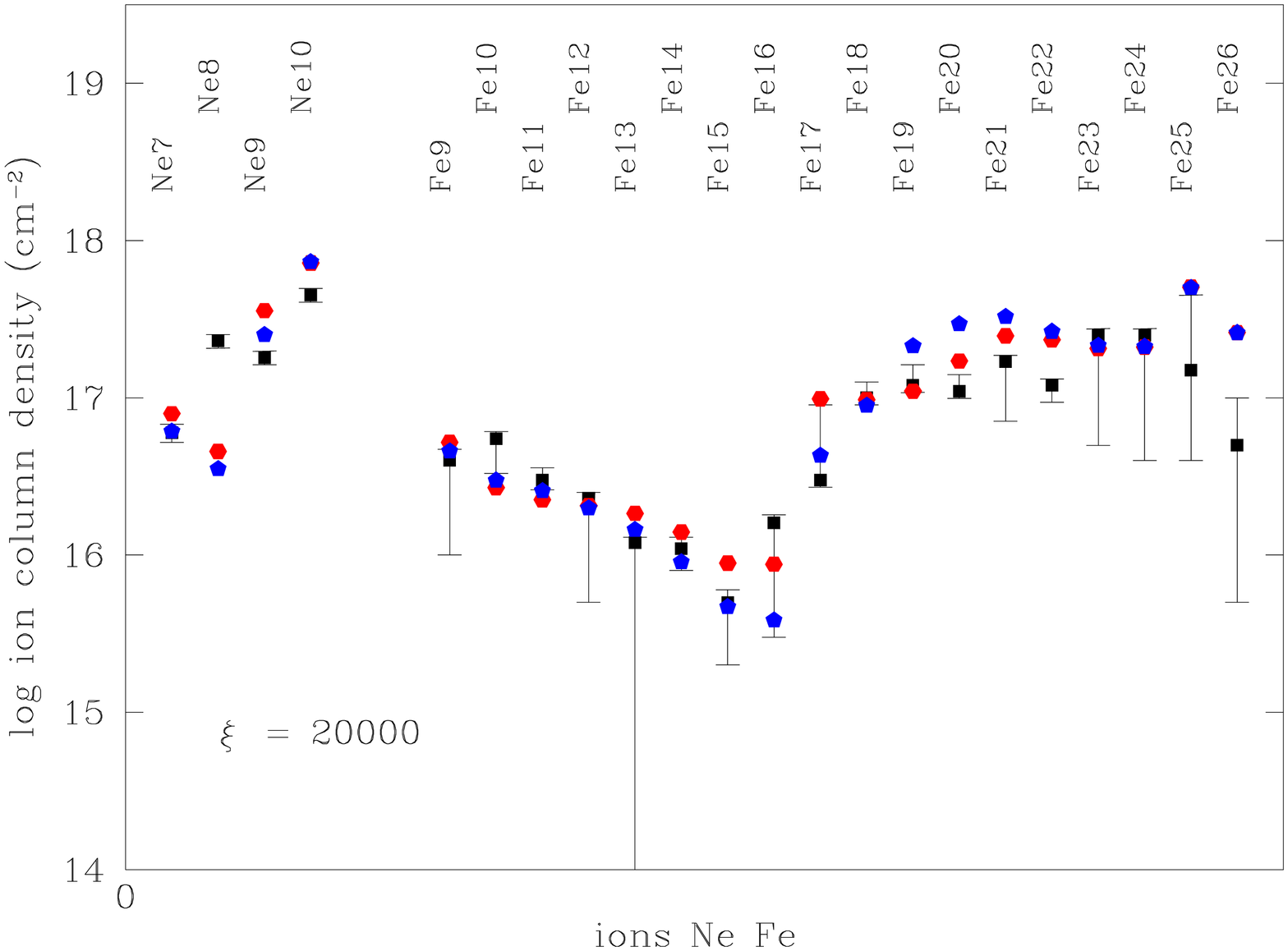}
  \caption{Observed column densities of individual ionic species,  $N_{\rm i,obs}$, in the warm absorber of NGC~3783 compared to the model results. The five figures correspond to different values of the ionization parameter, $\xi_{\rm tot}$, with their corresponding total hydrogen-equivalent column densities $N_{\rm H}^0$ (see Table 1). The observed data (black squares) carry error bars, the models are indicated by blue pentagons and red hexagons for the hot and cold solution, respectively. Top-left: $\xi_{\rm tot} = 2000$ ergs~cm~s$^{-1}$, Top-right: $\xi_{\rm tot} = 4000$ ergs~cm~s$^{-1}$, Middle-left: $\xi_{\rm tot} = 8000$ ergs~cm~s$^{-1}$, Middle-right: $\xi_{\rm tot} = 16000$ ergs~cm~s$^{-1}$, Bottom-left: $\xi_{\rm tot} = 20000$ ergs~cm~s$^{-1}$. Arabic numbers ion numbers are equivalent to Roman numerals.}
  \label{fig:titanNi-HolczerNi}
\end{figure*}

For each pair of model parameters ($\xi_{\rm tot}$, $N_{\rm H}^0$) best reproducing the shape of the observed continuum spectrum, we compare the ionic column densities derived from the measured equivalent widths (denoted $N_{\rm i,obs}$ hereafter and tabulated in Table~\ref{tab:Ni}) with the corresponding model values obtained by integrating the column density of each ionic species along the line of sight (denoted $N_{\rm i,mod}$). We concentrate on the ions of iron and neon spanning a wide range in ionization potential and used to build the observational AMD.

The values are reported in Fig.~\ref{fig:titanNi-HolczerNi}, for both cold and hot solutions. For the model at higher turbulent velocity, the ionic fractions differ on average by a factor  of $15\%$ from the values obtained with the nominative velocity value of 150 km/s.

\begin{figure*}[!t]
  \centering
  \includegraphics[width=0.98\textwidth,clip]{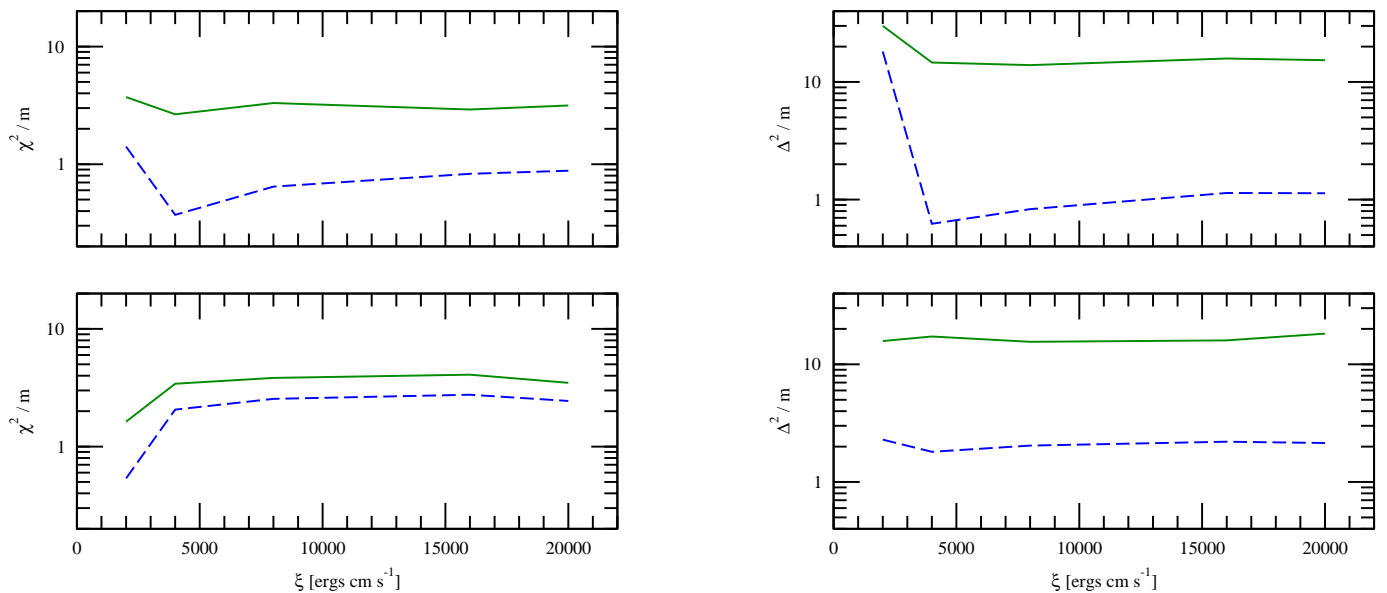}  
  \caption{Reduced differences $\chi^2/m$ (left) and $\Delta^2/m$ (right) between the observed ionic column densities and the model results (see equations~\ref{eqn:chi}~and~\ref{eqn:delta}) for different values of the ionization parameter $\xi_{\rm tot}$ and computation mode. The top panels denote the cold, and the bottom panels the hot solutions. The computations of $\chi^2$ and $\Delta^2$ include the ions Ne{~\scriptsize VII--X} and Fe{~\scriptsize IX--XXVI} (green solid line) and only Fe{~\scriptsize IX--XXVI} (blue dashed line). \label{fig:delta_xi}}
\end{figure*}

As a whole, the modeled values are roughly in agreement with the measured ones for $\xi_{\rm tot} > 2000$~ergs~cm~s$^{-1}$. Nonetheless, in  all cases a large discrepancy is found for the iron species Fe~{\scriptsize IV} -- Fe~{\scriptsize VIII} (not shown in Fig.~\ref{fig:titanNi-HolczerNi}). These are possibly due to discrepancies in the atomic data entering the observational analysis and the modeling. Table~\ref{tab:Ni} shows that the $N_{\rm i,obs}$ of the ions of Fe~{\scriptsize IV}--Fe~{\scriptsize VIII} are also poorly constrained observationally. Therefore, we exclude them from our analysis.

\begin{table}
 \centering
 \caption{Measured ionic column densities $N_{\rm i,obs}$  from the 900~ksec {\it Chandra} spectrum of NGC~3783 with upper and lower limits of the uncertainty, $N_{\rm i,obs}^+$ and $N_{\rm i,obs}^+$, respectively. All column densities are given in units of $10^{16}$~cm$^{-2}$.}
 \label{tab:Ni} 
 \begin{tabular}{lrrr}
  \hline\hline
     Ion  &  $N_{\rm i,obs}$  &  $N_{\rm i,obs}^+$  &  $N_{\rm i,obs}^-$\\
  \hline
     Ne{~\scriptsize VII}    &  6.0 &  6.8 &  5.2\\
     Ne{~\scriptsize VIII}   & 23.0 & 25.3 & 20.7\\
     Ne{~\scriptsize IX}     & 18.0 & 19.8 & 16.2\\
     Ne{~\scriptsize X}      & 45.0 & 49.5 & 40.5\\
     Fe{~\scriptsize IV}     &  0.5 &  3.7 &  0.2\\ 
     Fe{~\scriptsize V}      &  1.0 &  1.4 &  0.0\\ 
     Fe{~\scriptsize VI}     &  1.5 &  1.7 &  0.0\\ 
     Fe{~\scriptsize VII}    &  2.0 &  2.2 &  0.0\\ 
     Fe{~\scriptsize VIII}   &  3.0 &  3.3 &  1.3\\ 
     Fe{~\scriptsize IX}     &  4.0 &  4.7 &  1.0\\ 
     Fe{~\scriptsize X}      &  5.5 &  6.1 &  3.3\\ 
     Fe{~\scriptsize XI}     &  3.0 &  3.6 &  2.6\\ 
     Fe{~\scriptsize XII}    &  2.3 &  2.5 &  0.5\\ 
     Fe{~\scriptsize XIII}   &  1.2 &  1.3 &  0.0\\ 
     Fe{~\scriptsize XIV}    &  1.1 &  1.3 &  0.8\\ 
     Fe{~\scriptsize XV}     &  0.5 &  0.6 &  0.2\\ 
     Fe{~\scriptsize XVI}    &  1.6 &  1.8 &  0.3\\ 
     Fe{~\scriptsize XVII}   &  3.0 &  9.0 &  2.7\\ 
     Fe{~\scriptsize XVIII}  & 10.0 & 12.6 &  9.0\\
     Fe{~\scriptsize XIX}    & 12.0 & 16.2 & 10.8\\
     Fe{~\scriptsize XX}     & 11.0 & 14.0 &  9.9\\
     Fe{~\scriptsize XXI}    & 17.0 & 18.7 &  7.1\\
     Fe{~\scriptsize XXII}   & 12.0 & 13.2 &  9.4\\
     Fe{~\scriptsize XXIII}  & 25.0 & 27.5 &  5.0\\
     Fe{~\scriptsize XXIV}   & 25.0 & 27.5 &  4.0\\
     Fe{~\scriptsize XXV}    & 15.0 & 45.0 &  4.0\\ 
     Fe{~\scriptsize XXVI}   &  5.0 & 10.0 &  0.5\\
  \hline\hline
 \end{tabular}
\end{table}

For the lowest value of $\xi_{\rm tot} = 2000$~ergs~cm~s$^{-1}$, the model results differ significantly from the data. The offset between the observed and the modeled $N_{\rm i}$ is unacceptable for the ions Fe{~\scriptsize XXIII--XXVI}. At higher values of $\xi_{\rm tot}$, visual inspection of Fig.~\ref{fig:titanNi-HolczerNi} suggests a better agreement with the observations. We try to measure how accurately the model reproduces the observed column densities by calculating a reduced $\chi^2$ for each case:

\begin{equation}
  \chi^2 = \sum_{i=1}^m
           \frac{\left[ N_{\rm i,mod}   - N_{\rm i,obs}  \right]^2}
                {\left[ N_{\rm i,obs}^+ - N_{\rm i,obs}^- \right]^2},
  \label{eqn:chi}
\end{equation}

where $i$ runs through the total number $m$ of ionic species that are taken into account and $N_{\rm i,obs}^\pm$ are the upper and lower limits of the error bar attached to the observed value $N_{\rm i,obs}$ and reported in Table~\ref{tab:Ni}. The results for $\chi^2/m$ are plotted in Fig.~\ref{fig:delta_xi}~(left). The solid lines report $\chi^2/m$ computed for the cold (top) and hot (bottom) solutions including only the species of Fe{~\scriptsize IX--XXVI} and Ne{~\scriptsize VII--X}. This case corresponds to the series of ions chosen for the computation of the observational AMD presented in Sect.~\ref{sec:AMD_update}.

The comparison using $\chi^2/m$ is not entirely satisfactory: while $\chi^2/m$ adopts a minimum around $\xi_{\rm tot} = 4000$~ergs~cm~s$^{-1}$ for both sets of ions when assuming a cold solution, no minimum can be found in the case of the hot solution. The bottom-left panel of Fig.~\ref{fig:delta_xi} suggests that a minimum would require even smaller ionization parameters than $\xi_{\rm tot} = 2000$~ergs~cm~s$^{-1}$, which cannot be in agreement with the tendency read off the plots in Fig.~\ref{fig:titanNi-HolczerNi}. This contradictory behavior can be explained when examining the observational errors of $N_{\rm i,obs}$ in Table~2. These are particularly small for the ions Fe{~\scriptsize XVIII--XXII} giving these ions a strong impact when calculating $\chi^2/m$. In fact, for $\xi_{\rm tot} > 2000$~ergs~cm~s$^{-1}$, the column densities for Fe{~\scriptsize XVIII--XXII} are systematically overpredicted by the hot solutions. The ions Fe{~\scriptsize XXV} and Fe{~\scriptsize XXVI}, on the other hand, have large observational errors and their contribution to the calculation of $\chi^2/m$ is thus very small.

\begin{figure}[!th]
  \centering
  \includegraphics[width=0.48\textwidth,clip]{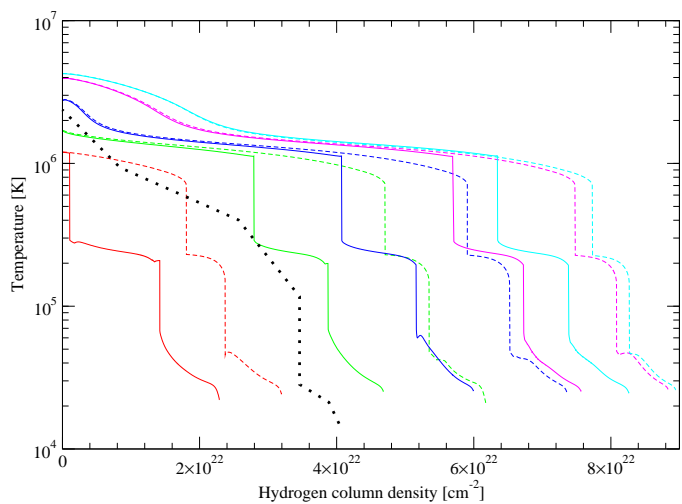}
  \caption{Temperature profiles as a function of the cumulative column density obtained from the {\sc titan} modeling of the warm absorber in NGC~3783. We adopt either the entirely cold (solid lines) or the entirely hot solution (dashed lines). From left to right, the lines represent the cases of $\xi_{\rm tot} = 2000$ ergs~cm~s$^{-1}$ (red), $\xi_{\rm tot} = 4000$ ergs~cm~s$^{-1}$ (green), $\xi_{\rm tot} = 8000$ ergs~cm~s$^{-1}$ (blue), $\xi_{\rm tot} = 16000$ ergs~cm~s$^{-1}$ (magenta) and $\xi_{\rm tot} = 20000$ ergs~cm~s$^{-1}$ (cyan). In each case, the value of $N_{\rm H}^0$ is adjusted to optimize the reproduction of the observed continuum spectral shape (see text). For comparison we also show the temperature profile obtained from the AMD analysis of the {\it Chandra} observation (black dotted line). \label{fig:T_profiles}}
\end{figure}

It is beyond the scope of this paper to provide a complete model grid that can be used systematically to fit observational data. This would require a much finer grid in $\xi_{\rm tot}$ and additional model parameters. Note, that in our approach we already fixed as many parameters as possible from the observational analysis. We therefore prefer to perform a less strict comparison and define a parameter $\Delta^2$ to estimate the deviation between the observed and the modeled values on a logarithmic scale: 

\begin{equation}
  \Delta^2 = \sum_{i=1}^m
             \frac{\left[ \log \frac{N_{\rm i,mod}}{N_{\rm i,obs}} \right]^2}
                  {\left[ \log \frac{N_{\rm i,obs}^\pm}{N_{\rm i,obs}} \right]^2}.
  \label{eqn:delta}
\end{equation}

Between $N_{\rm i,obs}^+$ and $N_{\rm i,obs}^-$, we choose the value with the larger difference $|N_{\rm i,obs}^\pm - N_{\rm i,obs}|$. The estimator $\Delta^2$ is a logarithmic version of the $\chi^2$ measure. The square of $\log{\frac{N_{\rm i,mod}}{N_{\rm i,obs}}}$ in the numerator evolves by the same amount whether $N_{\rm i,mod}$ and $N_{\rm i,obs}$ differ by a factor $a$ or $1/a$. The same holds true for the denominator when comparing $N_{\rm i,obs}^+$ or $N_{\rm i,obs}^-$ to the observed value of $N_{\rm i,obs}$. We obtain a reduced $\Delta^2/m \sim 1$ when the model differs by the same factor from the observed ionic column densities as the upper/lower observational limits.

We plot $\Delta^2/m$ as a function of $\xi_{\rm tot}$ in Fig.~\ref{fig:delta_xi}~(right) for the two series of ionic species considered in the comparison. Again, the largest differences between model and observations are found for the lowest ionization parameter $\xi_{\rm tot} = 2000$ ergs~cm~s$^{-1}$. The best agreement is obtained for $\xi_{\rm tot} = 4000$ ergs~cm~s$^{-1}$ for both hot and cold solutions. However, the values change by only $\sim 15\%$ when increasing $\xi_{\rm tot}$ up to 20000 ergs~cm~s$^{-1}$. If the sample used to compute the differences $\Delta^2$ is restricted to the ions Fe{~\scriptsize IX--XXVI} (dashed lines), the lowest difference is obtained for $\xi_{\rm tot} = 4000$ ergs~cm~s$^{-1}$.

In summary, this leads to a range of acceptable values for $\xi_{\rm tot}$ between 4000 and 8000~ergs~cm~s$^{-1}$, for hot as well as for cold solutions. However, the values of $\Delta^2/m$ or $\chi^2/m$ in this interval of $\xi_{\rm tot}$ are systematically lower for the cold solution than for the hot solution, which indicates that the cold phase predominates in the thermally unstable regime of the warm absorber.

\subsection{Temperature and pressure profiles along the line of sight}
\label{sec:comp_T}

The temperature profiles of our models are plotted in Fig.~\ref{fig:T_profiles}. They all show characteristic features that have been discussed in the previous work of \citet{goncalves2007}. Here, the spectral shape is more complex, but the modeling still leads to thermal instability effects that split the temperature distribution into three different regimes. The drop between the high temperature regime at the irradiated side to the middle plateau region is very sharp for all cases, the drop to the low temperature regime is slightly more shallow. Both drops correspond to very similar ranges of temperatures for all computed models: the first one is found at $2.8\times 10^5 - 1.1\times 10^6$~K and the second one at $5.6\times 10^4 - 2.0\times 10^5$~K for the cold solutions, and at $2.5\times 10^5 - 7.0\times 10^5$~K  and $5.0\times 10^4 - 1.8\times 10^5$~K for the hot solutions. The discontinuities appear in the same temperature ranges for the model with a higher turbulent velocity.

\begin{figure*}
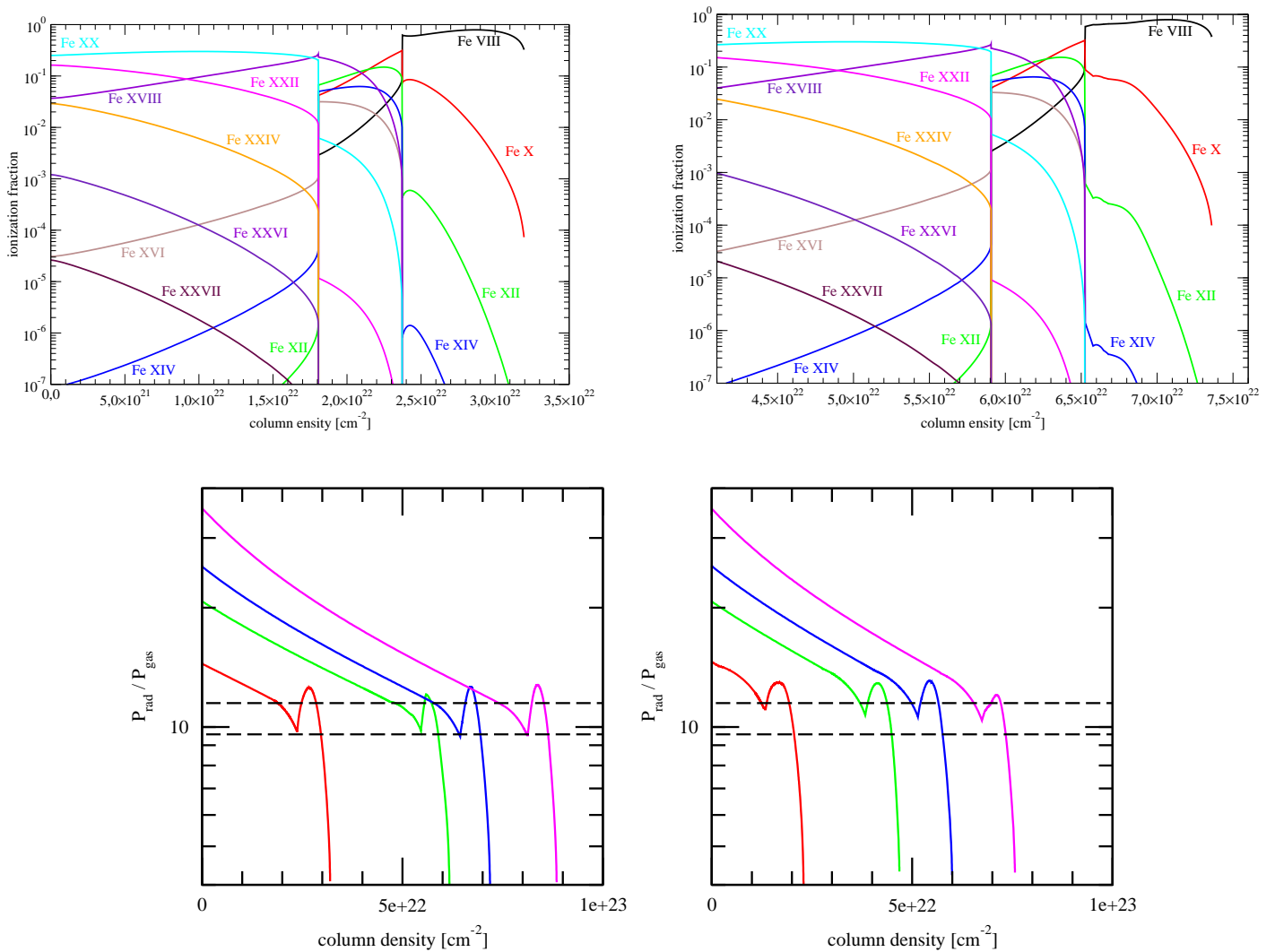

  \includegraphics[width=0.47\textwidth,clip]{figures/x2000c32e21As9_deg.eps}
  \hfill
  \includegraphics[width=0.47\textwidth,clip]{figures/x8000c738e20As9_deg.eps}\\
  \vfill
  \begin{center}
    \includegraphics[width=0.4\textwidth,clip]{figures/pressure_profiles_hot.eps}
    \includegraphics[width=0.4\textwidth,clip]{figures/pressure_profiles_cold.eps}
  \end{center}
  \caption{Fractional ion abundances of iron as a function of the hydrogen column density inside the warm absorber medium for the two hot modeling cases of $\xi_{\rm tot} = 2000$~ergs~cm~s$^{-1}$ (top-left) and $\xi_{\rm tot} = 8000$~ergs~cm~s$^{-1}$ (top-right). For $\xi_{\rm tot} = 8000$, the origin of the horizontal scale is at 4.1$\times 10^{22}$ for easier comparison with the case of $\xi_{\rm tot} = 4000$ plot (see text). The ratio between radiative and gas pressure for the uniformly hot (bottom-left) and cold (bottom-right) solutions are shown using the same color codes as in Fig.~\ref{fig:T_profiles}. The horizontal, dashed lines mark the pressure ratios at which the two thermal instabilities occur. \label{fig:ion_profiles}}
\end{figure*}

The strong resemblance of the temperature profiles for different parameters of $\xi_{\rm tot}$ is related to very similar profiles of fractional ionic abundances. In Fig.~\ref{fig:ion_profiles} (top), we compare the fractional abundance profiles for subsequent ionization states of iron between the two hot solutions of $\xi_{\rm tot} = 2000$~ergs~cm~s$^{-1}$ and $\xi_{\rm tot} = 8000$~ergs~cm~s$^{-1}$. The resulting profiles could practically be superimposed by applying a systematic shift in column density of $4.12 \times 10^{22}$~cm$^{-2}$. 

The modeling in \citet{goncalves2007} shows how the instability regime is related to a typical value of the ratio between radiative and gas pressure. We plot this ratio in Fig.~\ref{fig:ion_profiles} (bottom) for our models assuming uniformly hot and cold solutions. We note that the radiative pressure plotted here is derived from the spectral flux integrated over the complete wavelength range. The hydrogen column density of the medium remains moderate for all modeling cases representing less than 0.07 Thomson optical depths at maximum. Therefore, the pressure is dominated by the radiative pressure at the irradiated surface where the normalization of the profile rises in the order of increasing $\xi_{\rm tot}$. Going towards deeper layers of the medium, the upper dashed line in Fig.~\ref{fig:ion_profiles} (bottom) denotes the column density at which the first thermal instability and temperature drop occur. At these column densities, there is a slight kink in the profile of the pressure ratio. At the second occurrence of the thermal instability, when crossing the lower dashed line, an upturn of the profile can be seen. This behavior is very similar to the one described by \citet{goncalves2007} albeit for a different model setup. In the present work we find the high temperature instability at $P_{\rm rad}/P_{\rm gas} \sim 11.5$ for the uniformly hot and $P_{\rm rad}/P_{\rm gas} \sim 13.3$ for the uniformly cold solutions. The low temperature instability is triggered at $P_{\rm rad}/P_{\rm gas} \sim 9.6$ and $P_{\rm rad}/P_{\rm gas} \sim 10.6$ for the hot and cold solutions, respectively. For more details on the nature of the thermal instability and its relation to the net cooling functions we refer the reader to the analysis in \citet{goncalves2007}.

In contrast to the two instability drops in the theoretical profiles, the observed AMD in Fig.~\ref{fig:amd} only shows one ``forbidden range'' at $2.8\times 10^4 - 1.1 \times 10^5$~K. This zone can be associated to the second, low-temperature drop obtained in the modeling. The resolution of the observed AMD is not high enough to constrain the temperature profile more precisely, but the depression of the AMD at $T > 10^6$~K may be related to the high temperature drop seen in the models.

The maximum temperature at the irradiated side is about the same for both the cold and the hot solution of a given $\xi_{\rm tot}$, which is consistent with the fact that this region is thermally stable (see discussion below). A similar argument holds true for the lowest temperature values at the far side of the slab, where the medium is thermally stable as well. The maximum temperature of $2.4 \times 10^6$~K derived from the observed AMD is found between those obtained with the models of $\xi_{\rm tot} = 4000$ and $\xi_{\rm tot} = 8000$~ergs~cm~s$^{-1}$, while the minimum temperature value of $1.5 \times 10^4 $~K is slightly lower than those reached in all models. These low temperatures are constrained by the low ionization species such as Fe{~\scriptsize I--XVI}, whose lines constitute the so-called UTA features.

In Fig.~\ref{fig:T_profiles}, we also show the observational temperature profile obtained from the latest AMD analysis presented in Sect.~\ref{sec:AMD_update}. It suggests an ionization parameter slightly below $4000$~ergs~cm~s$^{-1}$, which is a bit low with respect to the result obtained from our comparison with the observed column densities in Sect.~\ref{sec:comp_Ni}. The discrepancy is less strong for the cold solution, which again seems to be the preferred thermal mode of the medium. 

We want to point out here, though, that the comparison of the observed temperature profile resulting from the AMD analysis with the modeling from {\sc titan} is not entirely straightforward. The observed temperature profile was obtained by fitting the absorption lines of various ions assuming only absorption but no reemission effects inside the medium. As described in \citet{holczer2007}, the fractional ionic abundances that are needed for the AMD analysis are computed by {\sc xstar} \citep{kallman2001} assuming a spherical geometry whereas here we apply {\sc titan} in plane-parallel geometry. Also the dependence of the ionic fractions with the ionization parameter is obtained assuming optically thin shells illuminated with the same incident spectrum, thus neglecting its distortion when passing through the medium. The differences in the two radiative transfer methods partly explain the discrepancies between the observational temperature profile and our modeling results.

\subsection{Comparison with the observed AMD}
\label{sec:AMD}

Finally, we attempt to reproduce the observed AMD of NGC~3783 given in Sect.~\ref{sec:AMD_update}. We construct a theoretical AMD from our modeling making use of the radiation to gas pressure ratio $\frac{P_{\rm rad}^{\rm ion}}{P_{\rm gas}}$ as a function of the cumulative hydrogen-equivalent column density $N_{\rm H}$ inside the medium. This ratio is related to the local ionization parameter $\xi$ and temperature $T$:

\begin{equation}
  \xi = 4\pi c k T \times 2.3 \; \frac{P_{\rm rad}^{\rm ion}}{P_{\rm gas}}
      = 1.2 \; {\rm T_4} \; \frac{P_{\rm rad}^{\rm ion}}{P_{\rm gas}},
  \label{eqn:amd_titan}
\end{equation}

with ${\rm T_4}$ being the local temperature in units of $10^4$~K. Equation (\ref{eqn:amd_titan}) is obtained from the relation between the two ionization parameters $\xi$ and $\Xi$ given in section~2a of \citet{krolik1981}. In contrast to the pressure ratio discussed in Sect.~\ref{sec:comp_T} and shown in Fig.~\ref{fig:ion_profiles} (bottom), the radiation pressure $P_{\rm rad}^{\rm ion}$ is based only on the ionizing flux between 1~and 1000~Ryd, and not integrated over the full spectral range. We point out that outside the interval of 1--1000~Ryd absorption processes are much less important than electron scattering. Therefore $P_{\rm rad}^{\rm ion}$ and $P_{\rm rad}$ evolve very similarly as a function of the medium temperature but have a different normalization. The values on the right hand side of equation (\ref{eqn:amd_titan}) are computed by {\sc titan} as a function of $N_{\rm H}$. In this way, the relation between $N_{\rm H}$ and $\log \xi$ can be established and then differentiated to construct a model AMD as defined in equation (\ref{eqn:amd}). 

\begin{figure*}[t]
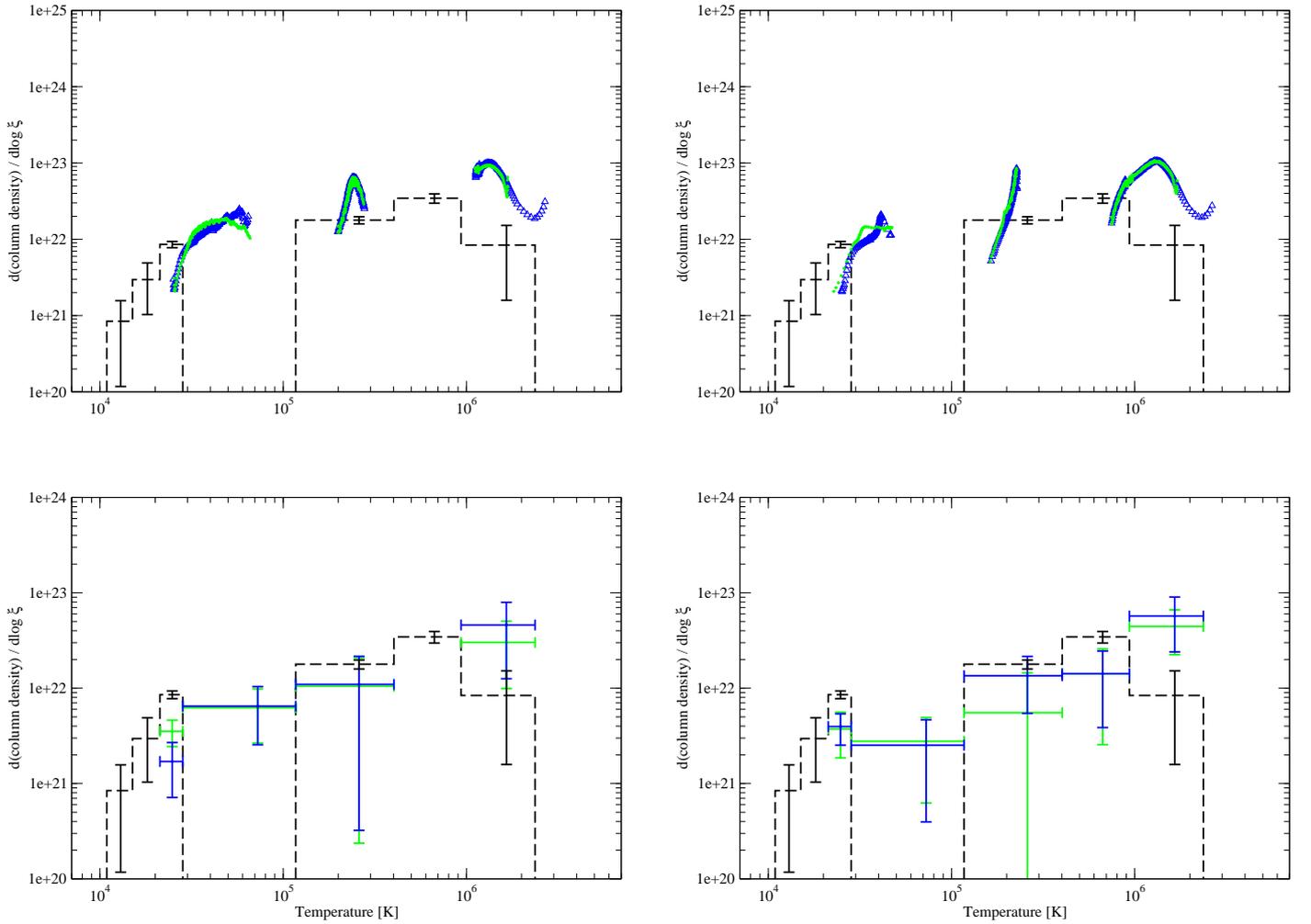

  \includegraphics[angle=0,width=0.48\textwidth]{figures/AMD_versus_logxi_logT_xspec_cold.eps}
  \hfill
  \includegraphics[angle=0,width=0.48\textwidth]{figures/AMD_versus_logxi_logT_xspec_hot.eps}\\~\\~\\
  \includegraphics[angle=0,width=0.48\textwidth]{figures/AMD_versus_logxi_logT_xspec_cut_rebinned_cold_scaleNH1e24.eps}
  \hfill
  \includegraphics[angle=0,width=0.48\textwidth]{figures/AMD_versus_logxi_logT_xspec_cut_rebinned_hot_scaleNH1e24.eps}  
\caption{Comparison between the observed and the modeled AMD as a function of temperature inside the medium (see Sect.~\ref{sec:AMD_update}). We construct theoretical AMD curves for the cold (left) and hot (right) solutions of the cases $\xi_{\rm tot} = 4000$ (green) and $\xi_{\rm tot} = 8000$ (blue). The observational AMD is denoted by the dashed line. The botom panels show the same theoretical AMDs as above but degraded to the resolution of the observed AMD and plotted on a larger vertical scale. \label{fig:amd_T}}
\end{figure*}

We prefer to plot the AMD as a function of the temperature rather than the ionization parameter. A relation between $T$ and $\xi$ is easily found because both values are computed as a function of $N_{\rm H}$ by {\sc titan}. This way we minimize possible systematic differences between the observed and theoretical AMD that are related to the calculation of the ionization parameter. In the observational analysis using {\sc xstar} a medium in spherical geometry at constant density is assumed and radiative processes are considered only along the radial direction. For the model computations with {\sc titan} we assume a plane-parallel atmosphere in pressure equilibrium and conduct full radiative transfer calculations for multiple directions. Furthermore, {\sc xstar} includes a lot more spectral lines in the radiative transfer than {\sc titan} does. Such systematics lead to differences in the profiles of $\xi(T)$ inside the medium. Rescaling the model AMD back to the temperature profile partly mitigates these effects. 

On the other hand, we mention that additional physical processes such as thermal conduction are not included in {\sc titan} while they may be present in the warm absorber and their effect might explain part of the discrepancy between the observational and the theoretical AMD. The whole cloud being in pressure equilibrium and moving uniformly, no strong dynamical shocks are expected to be formed, though. Turbulence might induce shocks, but all observed X-ray lines are reproduced with pure photoionization without any signature of collisional plasma. We also note that in the well-studied Seyfert-2 galaxy NGC~1068 the warm absorber medium is supposedly seen from the side and \citet{kinkhabwala2002} did not find a temperature structure that would indicate additinal shock heating.

We show the theoretical and observed AMDs in Fig.~\ref{fig:amd_T} (top) for the cold and hot solutions of $\xi_{\rm tot} = 4000$~ergs~cm~s$^{-1}$ (green points)  and $\xi_{\rm tot} = 8000$~ergs~cm~s$^{-1}$ (blue points), these models being favored when compared to the observed ionic column densities (see Sect.~\ref{sec:comp_Ni}). Although the resolution of the observational AMD is quite low, it is in rough agreement with the theoretical curves. The comparison of the normalization is more satisfactory at lower temperature. Observational and theoretical AMD both show a gap around 10$^5$~K but in the model AMD a second gap is detected between $3 \times 10^5$~K and 10$^6$~K. The two gaps correspond to the steep decrease in the temperature profiles discussed in Sect.~\ref{sec:comp_T}. 

The steep branch in the middle block of the AMD for the hot solution (Fig.~\ref{fig:amd_T}, right) is directly related to the rapid change in negative slope of the corresponding temperature profile (see Fig.5, dashed lines). The ``triangular'' shape seen in the middle block of the cold solution is associated to the changing sign of the corresponding temperature slope (see Fig.5, solid lines).

Also, compared to the observational AMD the low temperature instability strip found in the modeling is slightly shifted by a factor of 2 or less towards higher temperatures. This behavior is also visible in Fig. 5, as the vertical drop of the observational temperature profile (dashed line) is shifted downward compared to the temperature profiles for $\xi_{\rm tot} =4000$ and $\xi_{\rm tot} =8000$ models.

This may again be related to the different methods and assumptions for the radiative transfer applied in the observational and theoretical determinations of the AMD. For instance, we recall the difference between the local theoretical $\xi$ value derived from the pressure ratio and the ionization parameter used to construct the AMD which is computed based on the same spectral shape (the incident spectrum) whatever is the position of the associated dominant ion inside the cloud. Also physical processes like thermal conduction are not included in our model computations while they may induce changes by a few percent at the location of the discontinuities \citep{czerny2009}; this effect would not yet explain the complete shift in the horizontal direction.
 
The fact that the observational AMD does not show the high-temperature drop may be real or simply due to its limited resolution. The bottom panels in Fig.~\ref{fig:amd_T} show the theoretical AMD degraded to the same resolution as for the observed AMD. For the two cold solutions (left), the second gap is still present while the first one is no more visible as the observational gap is now filled. The agreement is better for the hot solutions (right panel), where for both ionization parameters the disappearance of the high temperature gap occurs but the low observational gap is partly filled up. Nonetheless, the theoretical AMDs reproduce the normalization of the observed ones to a large extent, although the two AMDs are built for different physical and geometrical configurations. They are thus complementary: the observational AMD gives a first idea of the presence of thermal instabilities in the cloud and thus justifies the use of sophisticated codes to model the observations under more suitable physical conditions than simple constant density models.

In parallel to the work presented here, thermal instability and the absorption measure distribution was modeled for Mrk~509 with {\sc titan} \citep{adhikari2015}. Despite a different spectral energy distribution that is injected into the warm absorber of Mrk~509, thermal instability ranges occur and are associated to forbidden temperature ranges of the AMD. The discussion in \citet{adhikari2015} about the impact of physical parameters and different numerical schemes applied to compute the radiative transfer extends the pioneering work by \citet{rozanska2006}. Drawing similar conclusions, these works are insightful and complementary to what is presented here.


\section{Discussion}
\label{sec:discuss}

The ionic column densities derived from the {\it Chandra} observation $N_{\rm i,obs}$ are in qualitative agreement with the theoretical values, $N_{\rm i,mod}$, obtained for different models except for the case of $\xi_{\rm tot} = 2000$~ergs~cm~s$^{-1}$ (see Fig.~\ref{fig:titanNi-HolczerNi}). A more quantitative comparison leads to a satisfying agreement for $\xi_{\rm tot}$ values between 4000 and 8000 ergs~cm~s$^{-1}$ with a preference for the cold over the hot solutions in the instability ranges. The models can account remarkably well for the ionic species at low and high ionization degrees. Differences are yet expected, since the $N_{\rm i,obs}$ values are affected by line emission partly filling up the absorption lines. Furthermore, the available atomic data keep evolving and are still subject to uncertainties which may affect the analysis \citep{kallman2007}. The ions from Fe{~\scriptsize IV} to Fe{~\scriptsize VIII} having a large uncertainty, they have been excluded from the comparison.

\subsection{Variability aspects of the warm absorber}

The {\it Chandra} mean spectrum is composed of six separate observations spanned over one year and a half. \Citet{netzer2003} have reported the softness ratio versus the 2--10~keV flux for each of the six pointings. Four observations show low softness ratios (0.06--0.14) and are characterized by a low flux level of 3--6$\times 10^{-11}$~erg~cm$^{-2}$~s$^{-1}$, while the two remaining observations show high softness ratios (0.14--0.23) with fluxes covering the range of 4--7.5$\times 10^{-11}$~erg~cm$^{-2}$~s$^{-1}$. The two flux ranges overlap, but according to \citet{netzer2003} the four low-softness observations can be associated to a low state (LS) and the two remaining to a high state (HS). For the following discussion we consider the same scheme bearing in mind that \citet{krongold2003} chose a different grouping to define high and low state spectra. The four low and soft state observations cover 666~ks and, as expected, the 900~ks average {\it Chandra} spectrum has a continuum shape (softness = 0.092) that is in better agreement with the LS than with the HS. The mean flux level over 2--10~\AA~changes at most by a factor of 2 between the six observations.

\citet{netzer2003} could not assign the two flux states to a change in luminosity and ionization parameter; an additional soft component is required to account for the high state spectrum. In the framework of the modeling presented here, the question arises whether the two states can be identified by the two cold and hot solutions of a model in pressure equilibrium that otherwise is characterized by the same ionization parameter and column density. Besides the transition between the two hot and cold solutions should occur on thermal timescales of the order of a few weeks/months \citep[see equation~4 in][]{chevallier2007}. This is indeed achievable with the parameter values considered in the modeling of \citet{goncalves2007} and \citet{chevallier2007}.

We have produced the two solution spectra for a set of ($\xi_{\rm tot}$, $N_{\rm H}^0$) values. The spectrum corresponding to the hot solution is systematically softer, as expected if low ionization lines such as the UTA are weaker or absent, but the flux below 10~\AA~is almost constant (varies by less than 10\%) and thus this can hardly account for a flux ratio HS/LS increasing regularly from 1.3 at 3~\AA~to 1.8 at 10~\AA~\citep[see Fig.9][]{netzer2003}. 

We also investigated the possibility of a change of $\xi_{\rm tot}$, keeping a constant $N_{\rm H}^0$ and choosing the cold (hot) solution for the higher (lower) $\xi_{\rm tot}$ value. Without investigating an exhaustive, time-consuming grid of models, we can find a couple of solutions. For $N_{\rm H}^0 = 6 \times 10^{22}$ cm$^{-2}$, the $\xi_{\rm tot} = 8000$~ergs~cm~s$^{-1}$ (cold) and $\xi_{\rm tot} = 11300$~ergs~cm~s$^{-1}$ (hot) solutions correspond to softness ratios of 0.10 and 0.17 in agreement with the two observed softness states and differing in flux by a factor 1.6 in the range 2-10~\AA, also in agreement with both states.

In conclusion, we find that the two states cannot be simply associated with the two extreme stable solutions present in an isobaric RPC wind. Assuming the same column density, an increase of the ionization parameter is also required to account for the high/soft state transition.

Shorter temporal variability within a few days has also been investigated in NGC~3783 \citep[see, e.g.,][]{markowitz2005, summons2007, reis2012}. Using broad energy range observations with {\it Suzaku}, \citet{reis2012} concluded that while the emission of the central region changes, the warm absorber does not vary with time or flux. This is consistent with the fact that the medium would respond to the incident flux change with a timescale of the thermal instability typically longer than a few days.

\subsection{Reproducing the observed AMD with an isobaric RPC model}

The theoretical temperature profiles we obtain exhibit two sharp drops (Fig.~\ref{fig:T_profiles}). Their extensions are very similar for all $\xi_{\rm tot}$ and whatever thermal solution chosen (hot or cold). Based on the ratio of radiation and gas pressure (see Sect.~\ref{sec:comp_T}), we constructed an AMD curve for the models and compared them to the observed one. The latter gives a total column density $N_{\rm H}^0$ and temperature ranges in rough agreement with those derived for the best model solutions. However, while the AMD level is similar for both curves, only one drop is apparent in the observed AMD and its position roughly agrees with the second (cooler) drop produced in the modeled AMD. The agreement is better for the hot solution.

According to equation (\ref{eqn:amd}), the AMD is the gradient of $N_{\rm H}$ with ionization parameter along the line-of-sight. In the instability strips this gradient falls to zero, but in between the instability regions the AMD measures how ``fast'' the column density falls off over a given range of ionization parameter or temperature where the medium is thermally stable. The successful reproduction of the observed AMD by our modeling thus gives a rather detailed insight into the thermal profile of the warm absorber.

There is no clear indication that the drop of the AMD at higher temperature is apparent in the data. The low resolution (7 points only) of the observed AMD may explain part of the discrepancy, as well as the different radiative transfer methods implemented in {\sc xstar} and in {\sc titan} as mentioned in Sect.~\ref{sec:comp_T}. Nevertheless, our combined observational and modeling work excludes  for NGC~3783 a so-called ``continuous AMD'' that corresponds to a homogeneous flow and is observed in, e.g., NGC~5548 \citep{steenbrugge2005}. In \citet{behar2009} the AMD of five AGN are reported including NGC~3783; all of them have a gap at intermediate $\xi$ values. A discrete AMD is also reported for NGC~7469 \citep{blustin2007}, MCG-6-30-15 \citep{holczer2010} and NGC~3516 \citep{holczer2012}. Again, in almost all cases a gap is detected, which is consistent with a multi-zone absorber.

\Citet{behar2009} fit the observed AMD profiles and derived the power law index $\alpha$ for a density profile of the wind on small scales, $n(\delta r) \propto \delta r^{-\alpha}$, or on large scales, $n(r) \propto r^{-\alpha}$. Herein, $n$ and $r$ denote the hydrogen-equivalent number density and the distance from the ionizing source, respectively. The approach relies on a self-similar density structure and the geometrical dilution of the irradiation flux. The slight rise of the observed AMD for NGC~3783 implies indexes of $\alpha = 0.78 \pm 0.07$ on small and $\alpha = 1.22 \pm 0.07$ on large scales. These results imply that the hydrogen number density in the wind falls off with distance from the irradiating source which is in agreement with MHD models where a dependence $n \propto r^{-1}$ results from the large scale profile of the magnetic field.

The situation changes for RPC winds. \Citet{stern2014a} and \citet{stern2014b} compute theoretical AMD profiles of a RPC warm absorber using the code {\sc cloudy} and show that for a broad range of model parameters the AMD has a flat distribution as observed. The density profile in RPC winds rises together with the gas pressure as a function of the distance to the ionizing source, which is what we also obtain in our modeling of NGC~3783 with {\sc titan}. This supports the hypothesis that the warm absorber is confined by radiation pressure.

With respect to the broad analysis of \citet{stern2014a}, \citet{stern2014b}, the present modeling of NGC~3783 assumed a more complex shape of the incident spectrum. This may play a role for the presence of thermal instability regimes. Still, in the modeling of isobaric RPC winds done by \citet{goncalves2007}, the incident power law is of a rather simple shape and produces thermal instabilities, while for various other incident spectra no instabilities are observed \citep[,e.g.,][see also Sect. 6.3]{chakravorty2012}. A factor may also be that in the present model the full radiative transfer was computed on relatively small spatial scales. 

Using {\sc cloudy} in the RCP hypothesis, \citet{stern2014b} do not always find temperature discontinuities in their models. The observational gap around $\xi=20$ present in the AMD of all 6 sources analyzed in their paper is indeed not reproduced by any of their models, while it is produced with {\sc titan}, which uses a fine adaptive grid to define more precisely the position and steepness of the discontinuities. The exact conditions for the instability to occur remain to be constrained and also require a comparison between different radiative transfer methods. Such a study is beyond the scope of this paper. Here, we only state that the warm absorber of NGC~3783 can be modeled as a thermally unstable RPC wind that is in agreement with the observed shape of the AMD.

\subsection{Constant density versus RPC winds}

In previous spectral analyses of the warm absorber in NGC~3783, a number of two \citep{krongold2003} or three clouds \citep{netzer2003} were invoked to reproduce the long-look {\it Chandra} observation. In both analyses the individual absorbers were found to be in pressure equilibrium. Here we demonstrate that a medium in pressure equilibrium, i.e., imposing a constant sum of gas and radiative pressure, reproduces the spectral shape and the ionic column densities measured for the warm absorber of NGC~3783. In Fig.~\ref{fig:ion_profiles} (bottom) we report the ratio of the radiative to gas pressure versus the integrated column density for the favored (cold and hot) solutions of $\xi_{\rm tot}$. One can derive from the Fig.~\ref{fig:ion_profiles} (bottom) showing $P_{\rm rad}/P_{\rm gas} = P_{\rm tot}/P_{\rm gas}-1$ that the radiative pressure dominates almost across the entire slab and is nearly constant except at the very far side where it drops sharply. This is in good agreement with the assumption of a RPC wind as discussed in \citet{baskin2014} and \citet{stern2014a,stern2014b}.

For all {\sc titan} models at pressure equilibrium obtained for this study, two regions of the slab turn out to be thermally unstable, which offers two possible temperature solutions, a hot and a cold one. We chose to keep the same kind of solution for all instability regions in the entire slab. This way we compute physical limits to the possible range of thermal states the medium can adopt. The sharp drops occurring twice in the temperature profiles are due to the appearance (disappearance) of an additional stable solution at left (right) of the net cooling function \citep[see][]{goncalves2007}. These drops are intrinsically linked to the existence of thermal instabilities. The work by \citet{hess1997} and \citet{chakravorty2008,chakravorty2009,chakravorty2012} has underlined the importance of the irradiating continuum shape, chemical abundance and the atomic data for the occurrence of thermal instabilities. For the continuum of NGC~3783, which corresponds to an optical-to-X-ray slope $\alpha_{\rm ox} \sim 1.3$ and $\alpha =0.77$ (we refer to these papers for the definition of these slopes), \citet{chakravorty2009} obtained an unstable region for $\xi$ being in the range of 50--1200 ergs~cm~s$^{-1}$, while \citet{krolik2001} found a stability curve with a near vertical rise and stable solutions for the same range of $\xi$ values. From these comparisons, we have again to conclude that the occurrence of thermal instability in a given model setup is very sensitive not only to the parameterization but also to the computational method adopted.

\section{Summary and conclusions}
\label{sec:conclude}

Following a combined observational and modeling approach we studied the warm absorber of the Seyfert galaxy NGC~3783. We simultaneously reproduce different observational characteristics by doing detailed radiative transfer modeling inside an isobaric RPC wind.

\begin{enumerate}

\item The continuum shape of the long 900\;ks {\it Chandra} spectrum can be correctly described using a photo-ionization model of a RPC medium at constant total (gas+radiation) pressure. We obtain a range of solutions for pairs of the ionization parameter $\xi_{\rm tot}$ and the total column density $N_{\rm H}^0$. These results extend previous analyses conducted in \citet{goncalves2006}, \citet{goncalves2007} and \citet{goosmann2009} to higher values of the incident ionization parameter, updated element abundances and additional spectral lines introduced in the {\sc titan} code.

\item Having fixed most input parameters of our modeling from various observational constraints, we reproduce the observed distribution of ionic column densities for a range of ionization parameters. The degeneracy in $\xi_{\rm tot}$ can by broken by a more quantitative analysis that indicates a range of $\xi_{\rm tot} = 4000-8000$~ergs~cm~s$^{-1}$. In the instability regime, the medium is more likely to adopt the cold rather than the hot phase.

\item We construct model AMDs for $\xi_{\rm tot} = 4000$~ergs~cm~s$^{-1}$ and $\xi_{\rm tot} = 8000$~ergs~cm~s$^{-1}$ that are in agreement with the overall shape of the observed AMD for both the hot and cold solutions. The model AMDs show two instability bands with the low-energy instability approximately agreeing with the one drop seen in the observed AMD. Note, however, that these temperature gaps are partly filled when plotting the theoretical AMDs using the same rough temperature sampling as for the observational AMD.

\end{enumerate}

This work gives further support to the hypothesis that the warm absorber in Seyfert galaxies is a RPC wind. Modeling with {\sc titan} the radiative transfer inside the warm absorber of NGC~3783, it was possible to reproduce a whole range of observational features simultaneously. This includes the presence of a thermally unstable regime seen in the observed AMD. As several AGN have exhibited at least one gap in their observed AMD (see Sec. 6.2), it is tantalizing to extend the present work to other objects in order to have a more physical description of their warm absorber. This has already been started in parallel work to ours presented by \citet{adhikari2015}. It is likely that the occurrence of the thermal instability under the condition of pressure balance depends not only on the assumed physics but also on the computational details of the adopted modeling method. The spatial scale on which the radiative transfer is solved may play a role, too. This needs to be explored in more detail in the future.

\begin{acknowledgements}
This work has been supported by the Israeli-French {\it ASTROPHYSICS} program and by the French GdR PCHE / PNHE. Ehud Behar received funding from the European Unions Horizon 2020 research and innovation programme under the Marie Sklodowska-Curie grant agreement no. 655324, and from the I-CORE program of the Planning and Budgeting Committee (grant number 1937/12). The authors are gratetful to Suzy Collin, Jonathan Stern and Agata R\'o\.zansk\'a for their helpful comments on this paper. Finally, we thank the anonymous referee for constructive remarks.
\end{acknowledgements}

\end{document}